\documentclass[aps,pra,preprint,groupedaddress,amsmath,amssymb]{revtex4-1}
\usepackage{bm}
\usepackage{graphicx}
\usepackage[usenames, dvipsnames]{color}

\newcommand*\diff{\mathop{}\!\mathrm{d}}

\begin{document}
\title{Thermodynamics of a Periodically Driven Qubit}%

\author{Brecht Donvil}%
\email{brecht.donvil@helsinki.fi}
\affiliation{Department of Mathematics and Statistics, University of Helsinki, P.O. Box 68, 00014 Helsinki, Finland}
\date{June 16, 2017}
\begin{abstract}
We aim to give a pedagogic presentation of the open system dynamics of a periodically driven qubit in contact with a temperature bath. We are specifically interested in the thermodynamics of the qubit. It is well known that by combining the Markovian approximation with Floquet theory it is possible to derive a stochastic Schr\"odinger equation in $\mathbb{C}^2$ for the state of the qubit. We follow here a different approach. We use Floquet theory to embed the time-non autonomous qubit dynamics into time-autonomous yet infinite dimensional dynamics. We refer to the resulting infinite dimensional system as the dressed-qubit. Using the Markovian approximation we derive the stochastic Schr\"odinger equation for the dressed-qubit. The advantage of our approach is that the jump operators are ladder operators of the Hamiltonian. This simplifies the formulation of the thermodynamics. We use the thermodynamics of the infinite dimensional system to recover the thermodynamical description for the driven qubit. We compare our results with the existing literature and recover the known results.
\end{abstract}
\maketitle

\section{Introduction}
In a recent paper \cite{PekCal} an experimental setup was proposed to perform a calorimetric measurement of work performed on a quantum system. The setup consists of a driven qubit in contact with a finite sized electron bath at a certain temperature. The electron bath acts as a calorimeter, its temperature changes due to interactions between the qubit and electrons. By monitoring the temperature one can indirectly track the evolution of the qubit.

The experimental setup of \cite{PekCal} was theoretically studied by \cite{Paolo} in the case of an adiabatic or weak drive. The authors of \cite{Paolo} derived a fluctuation relation and verified the first law of thermodynamics: the change of the internal energy of the qubit is equal to the work performed on the system minus the heat dissipated to the environment. Finally \cite{Paolo} showed how the thermodynamic relations recover the known expressions in the infinite calorimeter limit \cite{Breuer2}.

We want to extend the results of \cite{Paolo} beyond a weak or adiabatic treatment of the drive to a more general periodic drive. To do this it is essential to have a clear understanding of thermodynamics of a periodically driven qubit in contact with an infinite sized environment. This is what we aim for in the present paper. To develop a formalism well attuned to extend the work of \cite{Paolo} to a general periodic drive \cite{DonvilCal}. 

We do this by lifting the problem to an infinite dimensional Hilbert space. The reason for doing so is that we can use Floquet theory to embed the periodically driven qubit into a time-autonomous dynamics in this infinite dimensional Hilbert space. We refer to the infinite dimensional time-autonomous system as the dressed-qubit. For the dressed-qubit we can derive a stochastic Schr\"odinger equation, see e.g. \cite{openpaper}.
 The advantage of the infinite dimensional Hilbert space is that the jump operators are ladder operators of the Hamiltonian of the dressed-qubit. 
This fact simplifies the formulation of the thermodynamics. We show how to recover from the infinite dimensional level of description the thermodynamics for the driven qubit. We compare our results with earlier results in \cite{Cuetara, AlickiFloquet, Gasparinetti1,Gelbwaser} and we show that our approach is in agreement. We get the same heat currents and fluctuation relation.
From the fluctuation relation we obtain an expression for the path-wise entropy production. The entropy production is an important indicator for the design and control of efficient engines at the micro and nano-scales see e.g. the discussion in \cite{DechantHeat, CampisiJukka}
in the quantum case and \cite{PaoloKay} and references therein for the classical counterpart.

The study of periodically driven systems in contact with environments is an active field with a rich literature. Some recent works in relation to quantum nanodevices include \cite{Cuetara, AlickiFloquet, Szczygielski, Gasparinetti1,SzAl2015,LeAlKo2012,KoDiHa1997,HoKeKo2009,Gelbwaser} and in relation to quantum-computers \cite{AlickiMaster}. It is therefore worth to briefly discuss how to put the contribution of the present work in the context of the existing literature.

Reference \cite{BPfloquet} combined Floquet theory \cite{Shirley, Zeldovich1} and the Markov approximation to derive a stochastic Schr\"odinger equation in $\mathbb{C}^2$. The authors of \cite{BPfloquet} also noticed the correspondence between transitions occurring in driven qubit dynamics and those between dressed-atom states \cite{tan}. By dressed-atom we mean an atom interacting with a fully second quantised electromagnetic field. 
The infinite dimensional dressed-qubit is known to be equivalent with the dressed-atom picture in a certain limit \cite{Shirley, Swain1, Guerin}, hence the name dressed-qubit. We embed the stochastic dynamics of the driven qubit in the stochastic dynamics of the dressed-qubit. This confirms the interpretation made by \cite{BPfloquet} of the dynamics of the driven qubit in terms of atom-dressed states.

To the best of our knowledge the study of the thermodynamics of the periodically driven qubit in the existing literature is based on the analysis of the Lindblad-Gorini-Kossakowski-Sudarshan equation, see e.g. \cite{open, Rivas}, for the state operator acting on the $\mathbb{C}^2$ Hilbert space. See in particular \cite{Cuetara, AlickiFloquet, Gasparinetti1,LeAlKo2012,Gelbwaser}.
References \cite{Cuetara,Gasparinetti1} extend this approach to the counting statistics formalism. We show that the thermodynamic description of the driven qubit which we derive from the dressed-qubit gives the same fluctuation relation and heat currents as \cite{Cuetara, AlickiFloquet, Gelbwaser,Gasparinetti1}.

Reference \cite{LevyKosloff} thoroughly discusses local and global master equations for interacting subsystems separately interacting with a heat bath. It is instructive to characterize the Floquet approach in the language of \cite{LevyKosloff}. In the local approach one derives a master equation by neglecting the interaction between both subsystems. In the global approach one derives a master equation by solving the two subsystems without neglecting the interaction. For this reason Floquet theory is a form of a global approach. One first solves the dynamics of the qubit and the drive. 
The weak drive limit is a local approach. For strong driving this approach can result in violations of the second law \cite{GeKoSk95}.

The paper is structured as follows:
Section \ref{sec:Floquet} focusses on Floquet theory for a closed periodically driven qubit \cite{Shirley, Zeldovich1}. In particular, we recall how the time-non autonomous periodic quantum dynamics on a finite dimensional Hilbert space can be mapped on a time-autonomous infinite dimensional system \cite{Shirley, Swain1, Guerin}.

In Section \ref{sec:model} we introduce the model of the periodically driven qubit interacting with a infinite sized electron bath.

Section \ref{sec:SSE} consists of two parts. In Subsection \ref{subsec:sseq} we recall the results of \cite{BPfloquet} on the stochastic Schr\"odinger equation for the periodically driven qubit. Subsection \ref{subsec:qp} contains the core result of this paper. First we formulate a master equation for the dressed-qubit following the method of \cite{openpaper}. Then we show that the stochastic evolution of the periodically driven qubit can be embedded in the stochastic evolution of the dressed-qubit. The periodically driven qubit corresponds to the dressed-qubit with a specific set of initial conditions. This ensures that the heat dissipated to the environment in the case of the dressed-qubit and in the case of the driven qubit are equal.

The last part of this paper leisurely discusses the recovery of the thermodynamics of the periodically driven qubit from the thermodynamic relations of the infinite dimensional dressed-qubit.
In Sections \ref{sec:mast} and \ref{sec:ther} we recover fluctuation relation and heat currents by \cite{Gasparinetti1,Cuetara, AlickiFloquet}. In Section \ref{sec:mast} we derive a Pauli master equation for the populations in the Floquet states. Using the definition of Lebowitz and Spohn \cite{Lebowitz} we obtain an expression for the average entropy production. We also derive a fluctuation relation which holds for both the driven qubit and the dressed-qubit. In Section \ref{sec:ther} we study the thermodynamics of the dressed-qubit: we verify the first and second law of thermodynamics for the dressed-qubit and the driven qubit.

The example of a constant drive is discussed in Section \ref{sec:Const}. We also consider the weak drive limit for the dressed-qubit. We recover the thermodynamic relations derived in the weak drive regime.
In Section \ref{sec:Mon} we look at the monochromatic drive. We derive the stochastic Schr\"odinger equation, discuss the thermodynamics of this example and compare to earlier results by \citep{AlickiFloquet}.
\section{Floquet theory}\label{sec:Floquet}
In this Section we aim to give a brief overview on the Floquet approach for a closed periodically driven qubit, with an emphasis of what is needed in the rest of the paper. The qubit has a Hamiltonian $H_q(t)=H+H_d(t)$. The periodic part of the Hamiltonian $H_d(t)$ is called the drive. The time evolution of the two level system is found by solving the set of equations
\begin{equation}\label{eq:set}
\begin{cases}
(H_q(t)-i\hbar\partial_t)\psi(t)=0\\
\psi(0)=\psi_0,
\end{cases}
\end{equation}
were the Hamiltonian $H_q(t+T)=H_q(t)$ is periodic. 

By Floquet's theorem there exists a periodic matrix $P_{t+T,0}=P_{t,0}$ and a matrix $D$ such that the solution of the set of equations can be written into the form
\begin{equation}
\psi(t)=P_{t,0}e^{-iD t}\psi_0.
\end{equation}
References \cite{Shirley} and \cite{Zeldovich1} showed that the matrix $D$ is Hermitian. It can be diagonalised in an orthonormal basis $\nu_\pm$ with real eigenvalues $\epsilon_\pm$, which are also called quasi-energies. The eigenvectors evolve as
\begin{eqnarray}
\varphi_\pm(t)	&=P_{t,0}e^{-iD t}\nu_\pm\\
			&=e^{-i\epsilon_\pm t}\phi_\pm(t)\label{eq:ineed}.
\end{eqnarray}
We have defined the Floquet states $\phi_{\pm}(t)\equiv P_{t,0}\nu_\pm$, which are periodic. 
It is clear that we can add $n\hbar 2\pi/T$ to the quasi-energy in the last line and replace $\phi_\pm(t)$ by $\phi_{\pm,n}(t)\equiv e^{in (2\pi/T) t}\phi_\pm(t)$ and still get the same solution $\varphi_\pm(t)$. In this sense we are free to choose what we call $\epsilon_\pm$. In the rest of this paper we choose $\epsilon_\pm$ to be in the zeroth Brillouin zone $]-\pi/T,\pi/T]$. The operator $P_{t',t}$ is then the operator that evolves the Floquet states $\phi_\pm(t)$ to $\phi_\pm(t')$.

Note that the above is only an existence result. To find the Floquet states in practice one solves the problem
\begin{equation}\label{eq:floquet}
\begin{cases}
(H_q(t)-i\hbar\partial_t)\phi_{\pm,n}(t)=\left(\epsilon_\pm+n\frac{2\pi}{T}\right)\phi_{\pm,n}(t)\\
\phi_{\pm,n}(t+T)=\phi_{\pm,n}(t).
\end{cases}
\end{equation}
Then $\nu_\pm=\phi_{\pm}(0)$.
This problem is an eigenvalue equation in a larger space: our original Hilbert space $\mathbb{C}^2$ extended with the space of $T$-periodic functions $L^2[0,T]$. The scalar product is extended by
\begin{equation}\label{eq:scalar}
\langle f,g\rangle_{ L^2}=\frac{1}{T}\int_0^T \diff \tau \langle f(\tau),g(\tau)\rangle,
\end{equation}
for periodic functions $f$, $g\in \mathbb{C}^2\otimes L^2[0,T]$.
In the enlarged Hilbert space the Floquet states form an orthonormal basis.
We define the Floquet Hamiltonian $H_F(\tau)$ as
\begin{equation}\label{eq:floquetHam}
H_F(\tau)\equiv H_q(\tau)-i\hbar\partial_\tau,
\end{equation}
it is Hermitian for the scalar product $\langle\,.\, ,\,.\, \rangle_{L^2}$. The Floquet Hamiltonian is the energy operator of the infinite dimensional system we call the dressed-qubit.

The action of the periodic matrix $P_{t,0}$ on a vector $\psi\in\mathbb{C}^2$ can be expressed in terms of the Floquet states: $P_{t,0}\psi=\sum_{r=\pm}\phi_r(t)\langle \nu_r,\psi\rangle$.
The solution of the set of equations (\ref{eq:set}) is
\begin{align}
\psi(t) 	&=\sum_{r=\pm} e^{-i\epsilon_r t}\phi_r(t)\langle\nu_r,\psi_0\rangle\label{eq:soleq}\\
		&=\sum_{r=\pm}\sum_{n\in\mathbb{Z}} e^{-i(\epsilon_r +n\frac{2\pi}{T})t}\phi_{r,n}(t)\frac{1}{T}\int_0^T \diff s e^{in \frac{2\pi}{T}s} \langle\nu_r,\psi_0\rangle\\
		&=\sum_{r=\pm}\sum_{n\in\mathbb{Z}} e^{-i(\epsilon_r +n\frac{2\pi}{T})t}\phi_{r,n}(t)\frac{1}{T}\int_0^T \diff s \, e^{in \frac{2\pi}{T}s} \langle P_{s,0}\nu_r,P_{s,0}\psi_0\rangle\label{eq:soleq2}.
\end{align}
Going to the last line, we used the unitarity of $P_{s,0}$.

The dynamics in of the dressed-qubit in the space $\mathbb{C}^2\otimes L^2[0,T]$ are given by the set of equations
\begin{equation}\label{eq:set2}
\begin{cases}
(H_F(\tau)-i\hbar\partial_t)\Psi(t,\tau)=0\\
\Psi(0,\tau)=P_{\tau,0}\psi_0,
\end{cases}
\end{equation}
it is a one dimensional vector valued Schr\"odinger equation with periodic boundary conditions in the variable $\tau$ and $t$ is the time coordinate. The second equation lifts the initial state of the qubit $\psi_0$ to the larger space. In principle one could take any initial state, but this one reproduces the evolution of the driven qubit. $\Psi(t,\tau)$ is the coordinate representation of a time dependent vector $\Psi(t)\in \mathcal{H}\otimes L^2[0,T]$. The advantage of this larger space is that the Hamiltonian $H_F(\tau)$ is time-autonomous. The solution of the set of equations (\ref{eq:set2}) can be expressed in terms of the Floquet states, the energy eigenbasis of $H_F(\tau)$, in a straightforward way:
\begin{align}\label{eq:soleq3}
\Psi(t,\tau)=\sum_r\sum_{n\in\mathbb{Z}} e^{-i(\epsilon_r +n\frac{2\pi}{T})t}\phi_{r,n}(\tau)\frac{1}{T}\int_0^T \diff s\, e^{in \frac{2\pi}{T}s} \langle \phi_r(s),P_{s,0}\psi_0\rangle.
\end{align}
Comparing the solutions for the driven qubit and the dressed-qubit given by (\ref{eq:soleq}) and (\ref{eq:soleq3}) we see that both solutions are related by
\begin{equation}\label{eq:closedeq}
\Psi(t,\tau)=P_{\tau,t}\psi(t).
\end{equation}
The above relation holds between the driven qubit and the dressed-qubit initialised in the zeroth Brillouin zone. By multiplying the lift of the initial conditions \eqref{eq:set2} and the left hand side of \eqref{eq:closedeq} by $e^{in (2\pi/T) \tau}$, the relation holds for the dressed-qubit initialised in any single Brillouin zone.
Equation \eqref{eq:closedeq} implies the relation  $\psi(t)=\Psi(t,\tau)\big|_{\tau=t}$ \cite{Peskin1, Pfeifer1}.
\begin{figure}
\centering
\includegraphics[scale=0.25]{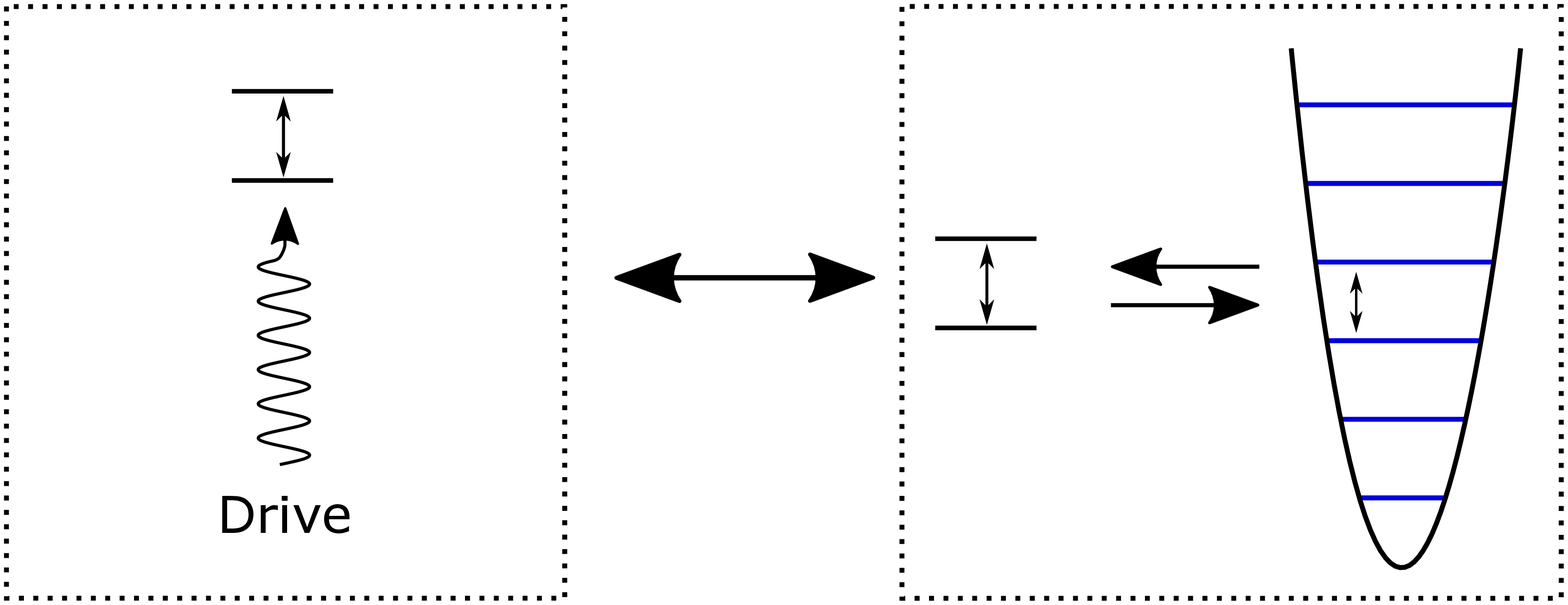}
\setlength{\unitlength}{0.1cm} 
\begin{picture}(0,0)
\put(-67,12.5){\scriptsize $H_d(t)$}
\put(-68,23.3){\scriptsize$\hbar\omega_q$}
\put(-31,15.7){\scriptsize$\hbar\omega_q$}
\put(-26,12){\scriptsize$H_{SP}$}
\put(-12,14.7){\scriptsize$\hbar\omega_L$}
\put(-90,-3.5){\scriptsize$H_F(\tau)=\frac{\hbar\omega_q}{2}\sigma_z+H_d(\tau)-i\hbar\partial_\tau$}
\put(-38,-3.5){\scriptsize$\frac{\hbar\omega_q}{2}\sigma_+\sigma_-+H_{SP}+\hbar\omega_La^\dagger a$}
\end{picture}
\caption{On the left side is the dressed-qubit with Hamiltonian $H_F(\tau)=\frac{\hbar\omega_q}{2}\sigma_z+H_d(\tau)-i\hbar\partial_\tau$, where $H_d(\tau)$ has period $2\pi/\omega_L$. In the appropriate limit \cite{Shirley,Zeldovich1,Guerin} it is equivalent to the system portrayed on the right: the atom interacting with the single mode of an electromagnetic field. $H_{SP}$ is the interaction between the system and photon corresponding to the drive $H_d(t)$.}
\label{fig:equiv}
\end{figure}

Moreover there is an established between the dressed-qubit system and the two level system interacting with a single mode of a fully second quantised electromagnetic field  with frequency $\omega_L= 2\pi/T$. In the limit of an intense laser field \cite{Shirley,Zeldovich1,Guerin} the Floquet states $\phi_{\pm,n}$ correspond to dressed-atom states with energy $\epsilon_\pm+n\hbar\omega_L$. The periodic phase factors $e^{in\omega_L \tau}$ can be interpreted as increasing the amount of photons by $n$ and $-\frac{i}{\omega_L}\partial_\tau$ is the number operator for the photons. The periodic drive $H_D(\tau)$ is the interaction between the photons and the atom. The periodically driven qubit is a semi-classical approximation of this interaction, where we completely forget about the interaction with the photons. The equivalence between the two systems is illustrated in Figure \ref{fig:equiv}.

\section{The model}\label{sec:model}
From now on we will focus on the actual system of interest: a periodically driven qubit in contact with a thermal electron bath.  The Hamiltonian of the full system consists of three terms
\begin{equation}\label{eq:hamiltonian1}
H(t)=H_q(t)+H_e+H_I.
\end{equation}
The first term on the right hand side is the Hamiltonian of the periodically driven qubit. It is time dependent and periodic with period $T$
\begin{equation}\label{eq:hamQ}
H_q(t)=\frac{\hbar\omega_q}{2}\sigma_z+H_d(t),
\end{equation}
where $\sigma_z$ is the canonical Pauli matrix and the time dependent term $H_d(t)$ represents the drive.  Let us define the angular frequency $\omega_L=\frac{2\pi}{T}$. The energy of the electrons is given by 
\begin{equation}\label{eq:hamE}
H_e=\sum_k \eta_k a^\dagger_k a_k,
\end{equation}
$a_k$ and $a^\dagger_k$ are ladder operators satisfying the fermionic anticommutation relations. The sum over $k$ is over all degrees of freedom for the electrons. The last term on the right hand side of equation (\ref{eq:hamiltonian1}) represents the interaction between the two systems
\begin{equation}\label{eq:hamI}
H_I=\sum_{kl}g_{kl}(\sigma_++\sigma_-)a^\dagger_ka_l.
\end{equation}
The diagonal elements of the interaction are assumed to be zero, i.e. $g_{kk}=0$ for all $k$ and $\sigma_\pm$ are ladder operators in the qubit space. Note that one could also be interested in coupling with a different thermal bath, e.g. a photon bath. The results derived in the following Sections do not depend on the specific thermal bath. It is only important that the jump rates satisfy detailed balance.

\section{Stochastic Schr\"odinger equation}\label{sec:SSE}
In this Section we discuss the evolution of a periodically driven qubit in contact with an electron bath. In Subsection \ref{subsec:sseq} we recall a result by \cite{BPfloquet} to obtain a stochastic Schr\"odinger equation in $\mathbb{C}^2$ for the state of the driven qubit.
In second Subsection \ref{subsec:qp} we derive a stochastic Schr\"odinger equation for the dressed-qubit system. We show how the stochastic evolution of the driven qubit can be embedded in the stochastic evolution of the dressed-qubit. 

\subsection{Periodically driven qubit}\label{subsec:sseq}
In \cite{BPfloquet} a stochastic Schrödinger equation is derived for a periodically driven system in contact with a thermal photon bath.
The approach can be extended in a straightforward way to systems in contact with a thermal electron bath with inverse temperature $\beta$. The only significant difference is in the jump rates. The exact form of the jump rates is given in Appendix \ref{sec:appSSE1}. For the results derived in the next Sections it is only important that the rates satisfy detailed balance.

The derivation by \cite{BPfloquet} assumes the existence of four timescales. The first timescale is $\tau_B$ over which the bath correlation functions decay. Secondly $\tau_R$ is the relaxation time of the open system, the qubit. $\tau_m$ is set by inverse of the dressed-qubit frequencies, i.e. the inverse of the difference in the quasi-energies $\epsilon_\pm/\hbar$ and $\omega_L$. The last timescale $\tau_T$ is the one over which the transitions are evaluated. It is assumed that $\tau_T$ relates to the others as $\tau_B,\tau_m\ll\tau_T\ll\tau_R$.

We formulate the stochastic Schr\"odinger equation in the interaction picture. A vector $\bar{\psi}$ is transformed to the interaction picture by setting $\psi=U^\dagger(t,0)\bar{\psi}$, where $\diff U^\dagger(t,0)/\diff t=iH_q(t)U^\dagger(t,0)/\hbar$. The stochastic Schr\"odinger equation describing the evolution of the qubit consists of two parts: a continuous evolution, proportional to $\diff t$, which is interrupted by sudden jumps
\begin{equation}\label{eq:SSE1}
\diff \psi(t)=-\frac{i}{\hbar}G(\psi(t)) \diff t +\sum_\omega \left(\frac{A(\omega)\psi(t)}{\|A(\omega)\psi(t)\|}-\psi(t)\right)\diff N(\omega).
\end{equation}
The operators $A(\omega)$ are called effect or jump operators, they determine which jumps the qubit can make. The operators are labelled by the amount of energy $\hbar \omega$ that is transferred from the qubit to the environment with the jump. They are expressed in terms of the Floquet states as \cite{BPfloquet}
\begin{eqnarray}\label{eq:effect1}
A(\omega)&=\sum_{r,s=\pm}\sum_n \alpha_{r,s,n} |\phi_{r}(0)\rangle\langle\phi_s(0)|
\end{eqnarray}
where $\phi_r$'s are the Floquet states, with the constraint on the sums for the energies
\begin{equation}\label{eq:energyconstraint}
\hbar\omega=\epsilon_s-\epsilon_r+n\hbar\omega_L
\end{equation}
and the matrix element $\alpha_{r,s,n}$ is defined by
\begin{equation}\label{eq:alpha}
\alpha_{r,s,n}=\frac{1}{T}\int_0^T \diff \tau\langle\phi_{r}(\tau),(\sigma_++\sigma_-)\phi_{s,n}(\tau)\rangle.
\end{equation}

The first term on the right hand side is proportional to the short time increment $\diff t$. It gives the continuous evolution of the wave function when no jumps occur
\begin{equation}\label{eq:contev3}
G(\psi)=-\frac{i\hbar}{2}\sum_\omega \gamma(\omega)[A^\dagger(\omega)A(\omega)-\|A(\omega)\psi\|^2]\psi.
\end{equation}

The rest of the terms on the right hand side describe the jump behaviour. They are proportional to increments of Poisson processes $\diff N(\omega)$. These increments are either 1, meaning the qubit  makes a jump, or 0, no jump occurs. The expectation value of the increments, conditioned that at a time $t$ the qubit is in a state $\psi$, is
\begin{equation}\label{eq:condexpfloq}
\mathbb{E}(\diff N(\omega)|\psi)=\gamma(\omega)\|A(\omega)\psi\|^2 \diff t,
\end{equation}
where $\gamma(\omega)$ is called the jump rate. The rates satisfy detailed balance
\begin{equation}\label{eq:detbal}
\frac{\gamma(\omega)}{\gamma(-\omega)}=e^{\beta\hbar \omega}.
\end{equation}

From the definition of the effect operators given by equation (\ref{eq:effect1}) one can see that it is possible for the qubit to exchange different amounts of energy while doing the same jump. For example: the qubit can jump from $\phi_{+,n}$ to $\phi_{-,0}$ exchanging the difference in quasi-energies $\epsilon_+-\epsilon_-$ with the environment plus $n\hbar\omega_L$ on top of that. Interpreting the Floquet states as dressed-atom states, as suggested by \cite{BPfloquet}, would explain this. The change in Brillouin zone means the creation or annihilation of photons with energy $\hbar\omega_L$. In the next section we derive a stochastic Schr\"odinger equation for the dressed-qubit. We embed the dynamics of the driven qubit in the dynamics of the dressed-qubit. In the end of Section \ref{sec:Floquet} we discussed the correspondence between dressed-qubit and dressed-atom states. This correspondence in combination with the embedding of the driven qubit in the dressed-qubit implies the interpretation proposed by \cite{BPfloquet}.

To show the above statements we make an extra assumption on the difference of the quasi-energies. We want to exclude the case in which the difference of the quasi-energies is a multiple of the photon energy. The physical consequence is that one can distinguish between a jump in the qubit state and a change in Brillouin zone, the creation of photons, by monitoring the energy exchange with the environment. We assume that
\begin{equation}\label{eq:assump}
\epsilon_+-\epsilon_-\neq k\hbar\omega_L, \quad \textrm{with} \quad k\in \mathbb{Z}.
\end{equation}
An equality in the above equation would correspond to a very specific choice of physical parameters. The difference in quasi-energies for the monochromatic drive, see Section \ref{sec:Mon}, is given by $\sqrt{\hbar^2(\omega_q-\omega_L)^2+4\lambda^2}$. For this to be equal to an multiple of $\hbar\omega_L$ a very non-generic choice of parameters is required.

Under assumption \eqref{eq:assump} there is a unique integer $n_\omega\in \mathbb{Z}$ such that the energy balance in equation (\ref{eq:energyconstraint}) holds. For each jump there is a unique amount of photons annihilated in the drive. The effect operators simplify to
\begin{eqnarray}\label{eq:effect2}
A(\omega)&= \sum_{r,s=\pm}\alpha_{r,s,n_\omega} |\phi_{r}(0)\rangle\langle\phi_s(0)|.
\end{eqnarray}

\subsection{Dressed-qubit}\label{subsec:qp}
Let us start by formulating a stochastic Scr\"odinger equation for the dressed-qubit. Remember that by dressed-qubit we mean the infinite dimensional system for which the closed dynamics are governed by the Floquet Hamiltonian $H_F$ \eqref{eq:floquetHam}.
The Hamiltonian for the dressed-qubit in contact with the electron bath is given by
\begin{equation}
H=H_{F}+H_I+H_E,
\end{equation}
where the different terms have been defined in equations (\ref{eq:floquetHam}), (\ref{eq:hamE}) and (\ref{eq:hamI}).
The dynamics are completely time-autonomous, the Hamiltonian does not explicitly depend on time anymore. Therefore a stochastic Schr\"odinger equation can be formulated by following the method described by \cite{openpaper}, this derivation requires the same timescales as discussed in the beginning of Subsection \ref{subsec:sseq}.  
We rewrite the interaction term in terms of energy eigenoperators, the jump operators, 
\begin{equation}
H_I=\sum_{kl}\sum_\omega g_{kl}B(\omega)a_k^\dagger a_l.
\end{equation}
The energy eigenoperators lower the energy of the system with $\hbar\omega$ and are defined by projecting different energy eigenstates onto $\sigma_++\sigma_-$. They act on a state $\Psi(t)$ as
\begin{eqnarray}\label{eq:effectph}
B(\omega)\Psi(t)=\sum_{r,s=\pm}\sum_{k,l}\phi_{r,k}\langle \phi_{r,k},(\sigma_++\sigma_-) \phi_{s,l}\rangle_{L^2}\langle\phi_{s,l},\Psi(t)\rangle_{L^2}
\end{eqnarray}
where the constraint (\ref{eq:energyconstraint}) on the energies holds with $n=l-k$. Comparing the above expression with expression (\ref{eq:effect1})  for the matrix elements of the effect operators $A(\omega)$, we see that  they are equal in the Floquet basis. The matrix element $\langle \phi_{r,k},(\sigma_++\sigma_-) \phi_{s,l}\rangle_{L^2}$ is equal to $\alpha_{r,s,l-k}$ as defined in equation \eqref{eq:alpha}. The effect operators determine which jumps a system can make. The qubit and the dressed-qubit can perform the same jumps exchanging the same amount of energy with the environment. 
Under assumption \eqref{eq:assump} the change in Brillouin zone $l-k=n_\omega$. The action of the effect operator on a state $\Psi(t)$ becomes
\begin{subequations}\label{eq:effectph2}
\begin{align}
\begin{split}
B(\omega)\Psi(t)=\sum_{r,s=\pm}\sum_{k} \alpha_{r,s,n_\omega}\phi_{r,k} \langle\phi_{s,k+n_\omega},\Psi(t)\rangle_{L^2}.
\end{split}
\end{align}
A more explicit version of this equality reads
\begin{align}
\begin{split}
(B(\omega)\Psi(t))(\tau)=\sum_{r,s=\pm}\sum_{k} \alpha_{r,s,n_\omega}\phi_{r,k}(\tau) \langle\phi_{s,k+n_\omega},\Psi(t)\rangle_{L^2}.
\end{split}
\end{align}
\end{subequations}
The stochastic Schr\"odinger equation for the dressed-qubit in contact with the electron bath is given by
\begin{eqnarray}\label{eq:sseph}
\diff \Psi(t)=-\frac{i}{\hbar}K(\Psi(t)) \diff t +\sum_\omega \left(\frac{B(\omega) \Psi(t)}{\|B(\omega) \Psi(t)\|_{L_2}}-\Psi(t)\right)\diff M(\omega).
\end{eqnarray}
The continuous evolution is defined analogous to last subsection
\begin{eqnarray}\label{eq:ssecontph}
K(\Psi)&=-\frac{i\hbar}{2}\sum_\omega \gamma(\omega)[B^\dagger(\omega) B(\omega)-\|B(\omega) \Psi\|_{ L_2}^2]\Psi
\end{eqnarray}
and the conditional average of the Poisson processes are
\begin{equation}\label{eq:condexpph}
\mathbb{E}(\diff M(\omega)|\Psi)=\gamma(\omega)\|B(\omega)\Psi\|_{ L_2}^2 \diff t.
\end{equation}
The jump rates $\gamma(\omega)$ are equal to those in equation \eqref{eq:condexpfloq}.

Let us now discuss the relation between the stochastic evolution for the periodically driven qubit and for the dressed-qubit, given by equations \eqref{eq:SSE1} and \eqref{eq:sseph} respectively. We have already shown that both systems can make similar jumps exchanging the same amount of energy and with the same jump rates. 
A mapping between the driven-qubit and dressed-qubit can be found if the dressed qubit is initialised as in \eqref{eq:set2}
\begin{equation}\label{eq:init}
\psi_0 \longleftrightarrow\Psi_0=P_{\tau,0}\psi_0= \phi_{+,0}(\tau) \langle \phi_+(0),\psi_0\rangle +\phi_{-,0}(\tau) \langle \phi_{-}(0),\psi_0\rangle
\end{equation}
In fact it does not matter in which Brillouin zone the dressed-qubit is initialised, as we show below, as long as the initial state is in only one of them.

Under assumption \eqref{eq:assump} we know at any time $t$ the amount $\mu(t)$ the quanta of $\hbar \omega_L$ gained by the environment, the total decrease (or increase when $\mu(t)<0$)  in Brillouin zone during the process. During the continuous evolution there is no change in Brillouin zone, while a jump $A(\omega)$ lowers it by $n_\omega$. We can write the evolution of the qubit state $\psi(t)$ and the decrease in Brillouin zone $\mu(t)$ by a couple of stochastic differential equations
\begin{equation}\label{eq:couple}
\begin{cases}
\diff \psi(t)=-\frac{i}{\hbar}G(\psi)+\sum\limits_{\omega} \left(\frac{A(\omega)\psi}{\|A(\omega)\psi\|}-\psi\right)\diff N(\omega)\\
\diff \mu(t)=\sum\limits_\omega n_\omega \diff N(\omega).
\end{cases}
\end{equation}
In the Appendix \ref{sec:appSSE2} we show that the dressed-qubit and the driven qubit state are connected by
\begin{equation}\label{eq:equivvv}
\Psi(t,\tau)=e^{-i\mu(t)\omega_L\tau}P_{\tau,0}\psi(t),
\end{equation}
the short time increment of the state $\Psi(t,\tau)$ satisfies the stochastic Schr\"odinger equation for the dressed-qubit \eqref{eq:sseph} with initial condition \eqref{eq:init}. Under this initial condition the Poisson processes $N(\omega)$ in \eqref{eq:SSE1} and $M(\omega)$ in \eqref{eq:sseph} are the same.
 We can initialise the dressed-qubit in an arbitrary Brillouin zone $m$ by replacing $\mu(t)$ by $\mu(t)-m$ in the above equation. 

In summary relation \eqref{eq:equivvv} shows how the stochastic evolution of the driven qubit can be embedded in the stochastic evolution of the dressed-qubit by introducing the photon counting process $\mu(t)$. In general the dressed-qubit shows a more complex behaviour than described by \eqref{eq:couple}, it can be initialised in multiple Brillouin zones. The equivalence discussed at the end of Section \ref{sec:Floquet} implies that the jumps between Floquet states and the change Brillouin zones for the periodically driven qubit can be interpreted as jumps between dressed-atom states and the creation or annihilation of photons. 

\begin{figure}
\centering 
\includegraphics[scale=0.35]{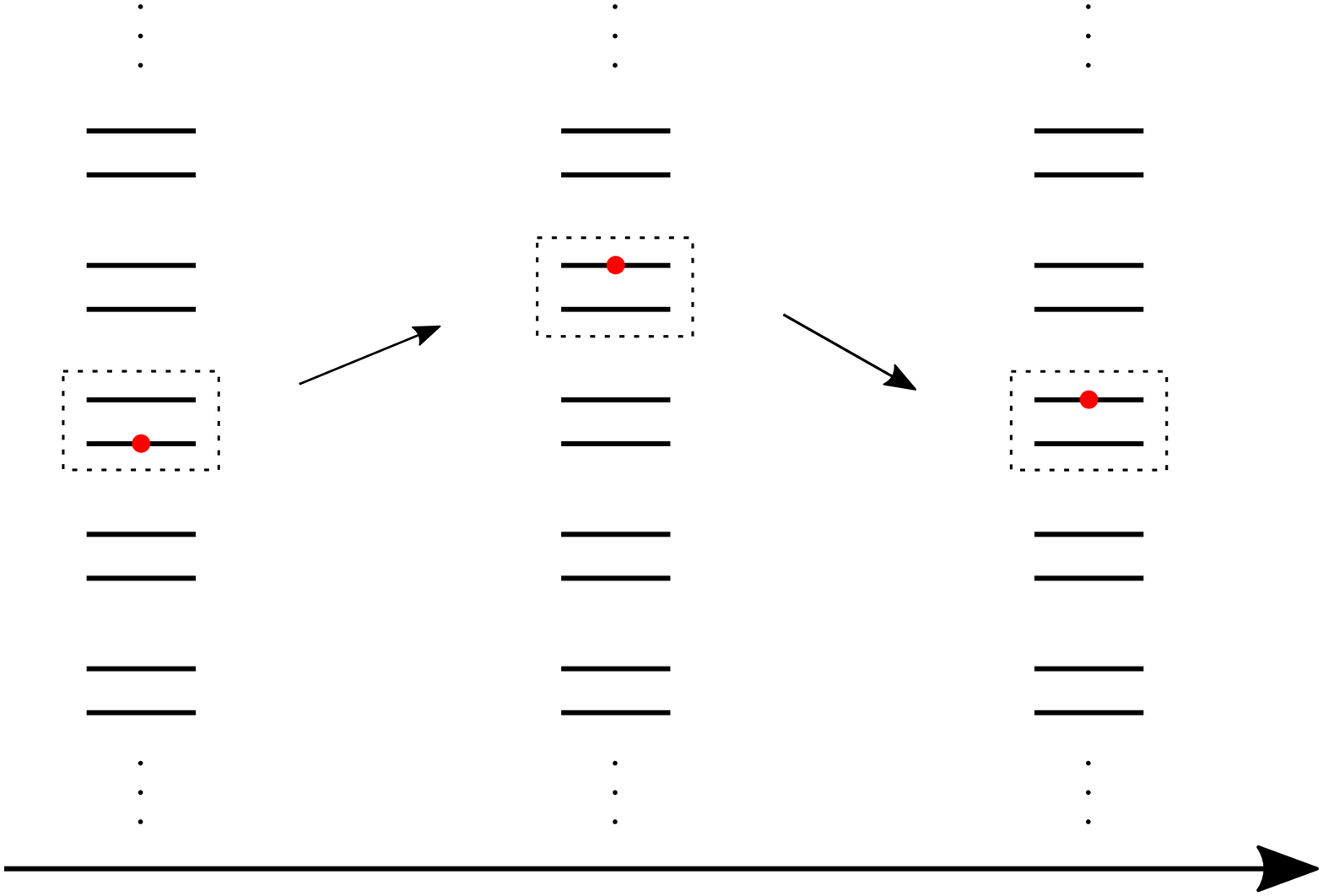}
\setlength{\unitlength}{0.1cm} 
\begin{picture}(0,0)
\put(-10,-4){$t$}
\put(-92,67){$\Psi(t_1)$}
\put(-88,-2){$t_1$}
\put(-90.5,38.5){\scriptsize$\psi(t_1)$}
\put(-95,52){$2$}
\put(-95,42){$1$}
\put(-109,33){$\mu(t_1)=$}
\put(-95,33){$0$}
\put(-98,23){$-1$}
\put(-98,13){$-2$}
\put(-58,67){$\Psi(t_2)$}
\put(-54,-2){$t_2$}
\put(-56.5,48){\scriptsize$\psi(t_2)$}
\put(-61,52){$2$}
\put(-61,42){$1$}
\put(-75,42){$\mu(t_2)=$}
\put(-61,33){$0$}
\put(-64,23){$-1$}
\put(-64,13){$-2$}
\put(-24,67){$\Psi(t_3)$}
\put(-20,-2){$t_3$}
\put(-22.5,38.5){\scriptsize$\psi(t_3)$}
\put(-27,52){$2$}
\put(-27,42){$1$}
\put(-41,33){$\mu(t_3)=$}
\put(-27,33){$0$}
\put(-30,23){$-1$}
\put(-30,13){$-2$}
\end{picture}
\caption{A visualisation of the stochastic process given by the couple of equations \eqref{eq:couple} and the process \eqref{eq:sseph} with jumps at times $t_1$, $t_2$, ... The qubit state is initialised in the lower Floquet state $\phi_-$ and initially no photons have been created or annihilated in the drive, so $\mu(t_0)=0$. At the jump times, the qubit can jump between different Floquet states and the photon number changes. }
\label{fig:fol}
\end{figure}

Finally, let us define the projector $P_n$ which projects $\Psi(t)$ onto the n-th Brillouin zone:
\begin{equation}\label{eq:projector}
 P_n \Psi(t)=\phi_{+,n}\langle\phi_{+,n}, \Psi(t)\rangle_{L^2}+\phi_{-,n}\langle\phi_{-,n}, \Psi(t)\rangle_{L^2}.
\end{equation}
The state $\Psi_n(t)= P_n \Psi(t)$ is living in the $n$-the Brillouin zone. From equation \eqref{eq:equivvv} we see that when the dressed-qubit is initialised as \eqref{eq:init}, all Brillouin zones are empty, i.e. $\Psi_n(t)=0$, except when $n=\mu(t)$. Figure \ref{fig:fol} shows a visualisation of the dressed-qubit process. 

\section{Master equation and entropy production}\label{sec:mast}
We calculate the average entropy production of the periodically driven qubit in two ways. In the next subsection we calculate it path-wise by comparing the probability measure of a realisation of the process and its backwards process. But first we start out by deriving a (classical) master equation for the Floquet states, a Pauli master equation, from the set of differential equations \eqref{eq:couple} With an equation like this we can use the classical definition for the entropy production rate by Lebowitz and Spohn \cite{Lebowitz}. 
\subsection{Master Equation}
Following \cite{openpaper} we define the state operator for the dressed-qubit system as $\rho(t)=\mathbb{E}(\Psi(t) \Psi^*(t))$. From the stochastic Schr\"odinger equation \eqref{eq:sseph} it is straightforward to check that the state operator satisfies the Lindblad-Gorini-Kossakowski-Sudarshan equation in the interaction picture
\begin{equation}\label{eq:lind}
\frac{\diff}{\diff t}\rho(t)=\sum_\omega \left( B(\omega)\rho(t)B^\dagger(\omega)-\frac{1}{2}\{B^\dagger(\omega)B(\omega),\rho(t)\}\right).
\end{equation}
Remember that $P_n$ as defined in equation \eqref{eq:projector} is the projector on the $n$-th Brillouin zone.
The diagonal blocks of the state operator $\rho_n(t)=P_n\rho(t)P_n$ corresponding to the $n$-th Brillouin zone satisfy a closed set of coupled (Lindblad) equations. Indeed projecting both sides of the above equation on the left and right with the projector $P_n$ we find, with de definition of the effect operators with \eqref{eq:effectph2}, that
\begin{equation}\label{eq:lindd}
\frac{\diff}{\diff t}\rho_n(t)=\sum_\omega \left( B(\omega)\rho_{n+n_\omega}(t)B^\dagger(\omega)-\frac{1}{2}\{B^\dagger(\omega)B(\omega),\rho_n(t)\}\right).
\end{equation}

We define $P(r,n,t)=\langle\phi_{r,n},\rho_n(t)\phi_{r,n}\rangle_{L_2}$ as the population in the state $\phi_{r,n}$. From equation \eqref{eq:lindd} we get a closed equation for the probabilities
\begin{equation}\label{eq:master}
\frac{\diff}{\diff t}P(r,n,t)=\sum_s\sum_m\left[ W_{m-n}(r|s)P(s,m,t)-W_{n-m}(s|r)P(r,n,t)\right].
\end{equation}
This equation is of the form of the master equation derived for dressed-atom states by Cohen-Tannoudji and Reynaud \cite{tan}.
The rates only depend on the difference between Brillouin zones and the Floquet states
\begin{equation}
W_{m-n}(r|s)=\gamma(\epsilon_s/\hbar-\epsilon_r/\hbar+(m-n)\omega_L)|\alpha_{r,s, m-n}|^2,
\end{equation}
where the matrix element $\alpha_{r,s, m-n}$ was defined in equation \eqref{eq:alpha}.
From the definition of Lebowitz and Spohn \cite{Lebowitz} we get the average entropy production rate corresponding to the master equation \eqref{eq:master}
\begin{equation}\label{eq:entprod}
\sigma(t) = \frac{1}{2}\sum_{r,s}\sum_{k,m}\left[\bigg( W_{n-m}(r|s)P(s,n,t)-W_{m-n}(s|r)P(r,m,t)\bigg)\log\left(\frac{W_{n-m}(r|s)P(s,n,t)}{W_{m-n}(s|r)P(r,m,t)}\right)\right].
\end{equation}
This is the average entropy production obtained by \cite{Cuetara}.

Summing over $n$ on both sides of equations \eqref{eq:lindd} we get a Lindblad equation for the state operator of the qubit
\begin{equation}\label{eq:linddd}
\frac{\diff}{\diff t}\bar{\rho}(t)=\sum_\omega \left( A(\omega)\bar{\rho}(t)(t)A^\dagger(\omega)-\frac{1}{2}\{A^\dagger(\omega)A(\omega),\bar{\rho}(t)\}\right),
\end{equation}
where $\bar{\rho}(t)=\sum_{n=-\infty}^{+\infty}\rho_n(t)$. It is straightforward to check that this is the Lindblad equation stemming from the stochastic Schr\"odinger equation for the driven qubit \eqref{eq:SSE1}.
The master equation for the populations $P(r)=\langle\phi_{r}(0),\bar{\rho}(t)\phi_{r}(0)\rangle=\sum_n P(r,n)$ can be obtained by taking the diagonal elements in the Floquet basis $\phi_\pm(0)$. The result is
\begin{equation}\label{eq:master2}
\frac{\diff}{\diff t}P(r,t)=\sum_s \left( W(r|s)P(s,t)-W(s|r)P(r,t)\right),
\end{equation}
where the rates $W(r|s)=\sum_n W_n(r|s)$ do not satisfy detailed balance any longer. The above equation can also be obtained by summing over $n$ on both sides of equation \eqref{eq:master}.

The entropy equation of Lebowitz and Spohn \cite{Lebowitz} still also applies to \eqref{eq:master2} and gives a positive definite entropy production $\bar{\sigma}(t)$.  Using the log sum inequality one can show that $\sigma(t)\geq \bar{\sigma}(t)$ \cite{Cuetara}.
\subsection{Pathwise entropy production}
We aim to calculate the path-wise entropy production for a realization of the stochastic process \eqref{eq:sseph} given that the dressed-qubit is initialised in a single Brillouin zone. Note that one could do the same derivation from the set of equations \eqref{eq:couple}, both approaches give the same result. We compare the probability density of a realization of the process, with the density of its time reversed process. We follow the same procedure as \cite{Paolo}.

Let us specify the measurement process. We wait until the dressed-qubit makes a jump at a time $t_i$ such that it is certainly in one of the Floquet states $\phi_{\pm,n}$, to start the measurement.  In an experiment one can take $t_i$ to be very large compared to the time scale of the evolution of the qubit such that it has almost certainly made a jump to one of the Floquet states. The Floquet states are stationary states for the continuous evolution of the stochastic Schr\"odinger equation \eqref{eq:ssecontph}. We are dealing with a pure jump process between the Floquet states.

The dressed-qubit has an initial probability distribution
\begin{equation}
P_i=P(i,n,t_i),
\end{equation}
where $i=\pm$ and $n\in\mathbb{Z}$ are the quantum specifying the initial state $\phi_{i,n}$.

At the end of the measurement $t_f$ the dressed-qubit is in the state $\phi_{f,m}$. We get a final distribution
\begin{equation}
P_f=P(f,m,t_f).
\end{equation}
In between the initial and final time a realisation is characterised by a set of times $t_i=t_1<...<t_n<t_{n+1}=t_f$ and jumps $B(\omega_i)$, which we denote as the set $\{t_j, B(\omega_j)\}_{j=2}^n$. In between the jumps the qubit stays in its current Floquet state.
The probability density of this realisation is given by \cite{open}
\begin{equation}\label{eq:probfl}
\mathbb{P}(\{t_j, B(\omega_j)\}_j,\phi_{f,m}|\phi_{i,n})=\prod_{j=2}^n \gamma(\omega_j) |\langle\phi_{f,m}, \prod_{k=2}^n B(\omega_k) \phi_{i,n}\rangle_{L^2}|^2.
\end{equation}
The reverse process is characterised by the initial state $\phi_{f,m}$, the final state $\phi_{i,n}$ and the set of jump times and jumps $\{t_f-t_{n-j}, B^\dagger (\omega_{n-j})\}_j$. The reversed process thus has probability density
$\mathbb{P}(\{t_f-t_{n-j}, B^\dagger (\omega_{n-j})\}_j,\phi_{i,n}|\phi_{f,m})$.

Comparing the probability of a realisation and its time reversed equivalent we get a fluctuation relation
\begin{equation}\label{eq:fluct}
\mathbb{P}(\{t_j, B(\omega_j)\},\phi_{f,m}|\phi_{i,n})=e^{ J}\mathbb{P}(\{t_f-t_j, B^\dagger (\omega_{n-j})\},\phi_{i,n}|\phi_{f,m}),
\end{equation}
where $J$ is the inverse temperature $\beta$ times the total heat dissipated to the environment
\begin{equation}
J=\beta\sum_k \hbar \omega_k.
\end{equation}

Combining the all contributions to the entropy production, we arrive at the total pathwise entropy production
\begin{equation}\label{eq:entpath}
S=S_f-S_i + \beta \sum_k \hbar \omega_k.
\end{equation}
The first two terms on the right hand side are the entropy of the initial preparation and the final measurement $S_{i/f}=-\log P_{i/f}$.
The third term is the inverse temperature $\beta$ times the heat dissipated from the dressed-qubit to the electron bath, due to jumps. The above equation recovers the result by \cite{Cuetara}. 

The instantaneous average entropy production is given by \cite{MaesEnt}
\begin{eqnarray}
\sigma&=\lim_{t\downarrow 0}\frac{1}{t}\langle  S\rangle,
\end{eqnarray}
 where $t$ is the length of the time interval for which we consider the entropy production. This corresponds to equation \eqref{eq:entprod}.

\section{Thermodynamics}\label{sec:ther}
In this section we discuss the thermodynamics of the dressed-qubit and the driven qubit. More precisely, we will use the thermodynamics relations obtained for the dressed-qubit to formulate those for the driven qubit.

Let us examine the thermodynamics of the dressed-qubit. From now on we will be working in the interaction picture. This means the $H_F$ is added to the continuous evolution \eqref{eq:ssecontph} of the stochastic process \eqref{eq:sseph}. We suppose that the system is in a state $\Psi(t)$ at time $t$. By construction the Hamiltonian $H_F$ is time-independent. Thus we expect the work performed on the infinite dimensional system to be zero. If we apply the definition of work by \cite{Pusz, AlickiHeatEngine} we find
\begin{equation}\label{eq:work}
\mathcal{W}(t)=\langle \Psi(t) ,\left(\frac{\diff }{\diff t}H_{F}\right) \Psi(t)\rangle_{L^2}=0.
\end{equation}
We emphasise that in the above equation $t$ is the time parameter. In this sense $H_F$ is time-independent, but it does have a periodic dependence on the variable $\tau$. The $L^2$ scalar product appearing in \eqref{eq:work} is the integral over $\tau$.
The heat $\mathcal{Q}$ dissipated from the dressed-qubit system to the electron bath at time $t$ is
\begin{equation}\label{eq:heat}
\mathcal{Q}= \sum_\omega \hbar \omega \diff M_\omega,
\end{equation}
where $\diff M_\omega$ are the Poisson processes appearing in the stochastic Schr\"odinger equation \eqref{eq:sseph}. The embedding of the driven qubit in the dressed-qubit \eqref{eq:equivvv} tells us that the above expression is also the heat dissipated by the driven qubit. 
The internal energy of the system is given by
\begin{equation}\label{eq:internalE}
\mathcal{E}(t)=\langle \Psi(t) ,H_{F}\Psi(t)\rangle_{L^2}.
\end{equation}
We retrieve the first law of thermodynamics in the form
\begin{equation}\label{eq:firstlaw1}
\mathbb{E}(\diff \mathcal{E}(t) |\Psi(t))=-\mathbb{E}( \mathcal{Q}|\Psi(t)),
\end{equation}
where $\mathbb{E}(.|\Psi(t))$ is the average over all realisations of the stochastic process \eqref{eq:sseph} conditioned on the state $\Psi(t)$.
The change in energy of dressed-qubit is equal to minus the heat dissipated to the environment. The second law of thermodynamics follows from equation \eqref{eq:entpath}. We have 
\begin{equation}
S_f-S_i=S-\beta \mathcal{Q}.
\end{equation}
The average of the entropy production S is positive by definition \eqref{eq:fluct}. Taking the average over all realisations of the stochastic process \eqref{eq:sseph}, we get
\begin{equation}\label{eq:secondlaw}
\mathbb{E}(S_f-S_i) \geq -\beta\, \mathbb{E}(\mathcal{Q}).
\end{equation}
This finishes the description of the thermodynamics for the dressed-qubit. Our aim is now to reinterpret these results in terms of quantities defined in the finite dimensional space of the driven qubit.

The identity $H_F(\tau)=H_Q(\tau)-i\hbar\partial_\tau$ implies that we can write the internal energy of the qubit \eqref{eq:internalE} as the sum of two contributions. We interpret
\begin{equation}
E(t)=\langle \Psi(t) ,H_Q\Psi(t)\rangle_{L^2} 
\end{equation} 
as the internal energy of the qubit. The differential of the second contribution
\begin{equation}\label{eq:workwork}
W(t)=-\diff\langle \Psi(t) ,-i\hbar\partial_\tau\Psi(t)\rangle_{L^2}.
\end{equation}
we interpret as the work performed by the drive on the qubit.
It follows immediately from \eqref{eq:firstlaw1} that 
\begin{equation}\label{eq:firstlaw2}
\mathbb{E}(\diff E(t) |\Psi(t))=\mathbb{E}(W(t)|\Psi(t))\diff t-\mathbb{E}( \mathcal{Q}|\Psi(t)).
\end{equation}
This is the energy balance relation from the qubit, the first law of thermodynamics for the qubit.
The second law of thermodynamics \eqref{eq:secondlaw} derived for the dressed-qubit also holds for the driven qubit.

\begin{figure}
\centering
\includegraphics[scale=0.2]{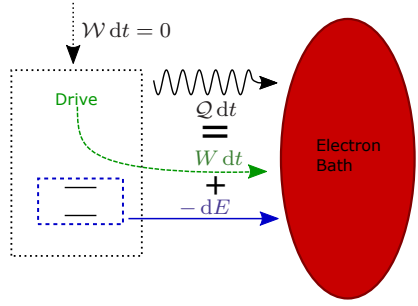}
\setlength{\unitlength}{0.1cm} 
\begin{picture}(0,0)
\put(-32,24.5){\scriptsize$\mathcal{Q}\diff t$}
\put(-32,18){\scriptsize \textcolor{OliveGreen}{$W\diff t$}}
\put(-34,11.5){\scriptsize \textcolor{BlueViolet}{$-\diff E$}}
\put(-47,35){\scriptsize$\mathcal{W}\diff t=0$}
\end{picture}
\caption{A schematic representation of the energy flows between the different subsystems for the two different levels of description: the dressed-qubit and the driven qubit. The curly (black) arrow shows the energy dissipated to the environment. The (black) dotted arrow is the work performed on the dressed-qubit, which is zero. 
For the the driven qubit we identify two contributions to the heat.
The normal (blue) arrow shows qubit component of the heat flow to the environment, i.e. the change in internal energy of the qubit $\diff E$.  The (green) dashed line is the drive contribution to the heat, the work performed by the drive.}
\label{fig:energy}
\end{figure}

Let us now examine the above equation when the qubit reaches its steady state. By this we mean the stationary solution of equation \eqref{eq:linddd}. The stationary solution is
\begin{equation}
\rho_s=\frac{1}{\Gamma_++\Gamma_-}\begin{pmatrix}
\Gamma_+ & 0\\0&\Gamma_-
\end{pmatrix},
\end{equation}
where $\Gamma_\pm=\sum_{n}\gamma(\epsilon_\mp-\epsilon_\pm +n\hbar\omega_L)$. 
A straightforward calculation shows that equation \eqref{eq:firstlaw2} becomes
\begin{align}\label{eq:statn}
\mathbb{E}(\diff W(t)|\rho_s)	&=-\hbar\omega_L\mathbb{E}\left(\sum_\omega n_\omega\diff N(\omega)\bigg|\rho_s\right)\\
								&=-\hbar\omega_L\mathbb{E}\left(\diff \mu(t)|\rho_s\right).
\end{align}
Remember that $\mu(t)$ is the process introduced in Subsection \ref{subsec:qp} to keep track of the amount of photons annihilated in the drive.
With equations \eqref{eq:firstlaw2} and \eqref{eq:statn} we recover the heat current by \cite{Gasparinetti1,Cuetara} and the work at steady state by \cite{Cuetara}. Figure \ref{fig:energy} shows a schematic representation of the energy flows on the two different levels of description: the dressed-qubit and the driven qubit.

The above equations show us how heat is not only dissipated to the environment by the qubit but also by the drive, due to the indirect coupling of the drive to the environment. In the limit that the strength of the drive goes to zero, the heat dissipated to the environment is equal to the heat dissipated by the qubit. Moreover they are equal to the heat dissipated to the environment when the stochastic Schr\"odinger equation is derived in the weak drive limit \cite{Breuer2}. In this limit the thermodynamics of both approaches are equivalent, it is sufficient to work in $\mathbb{C}^2$. This will be shown in Section \ref{sec:Const}.

\section{Constant drive}\label{sec:Const}
It is instructive to look at the case of a constant drive for two reasons. The first is that it allows a complete analytical treatment. The second is because we need it to recover from the infinite dimensional space the normal weak drive case as discussed at the end of Section \ref{sec:ther}.
The time independent Hamiltonian of the qubit with constant drive modulo a unitary transformation is given by
\begin{equation}\label{eq:timeind}
H_Q=\frac{\hbar\omega}{2}\sigma_z.
\end{equation}
The explicit solution of the spectral problem \eqref{eq:floquet} is given by the set $\{\pm\hbar\omega/2+n\hbar\omega_L\}_{n\in\mathbb{Z}}$ of eigenvalues and eigenvectors $\phi_{\pm,n}(\tau)=e^{in\omega_L\tau}|\pm\rangle$. The vectors $|\pm\rangle$ are the energy eigenstates of the time independent Hamiltonian $H_Q$. 

The only possible jumps for the dressed-qubit system are those exchanging no photons. The matrix elements of the jump operators \eqref{eq:alpha} are zero when photons are being created or annihilated. The jump operators are $|\pm\rangle\langle\mp|\otimes \mathbb{I}$, where $\mathbb{I}$ is the identity operator in $L^2$. We observe that the jumps are purely between the qubit energy states $|\pm\rangle$. 

The results from Sections \ref{sec:mast} and \ref{sec:ther} hold, with all transitions creating or annihilating photons set to zero. The work performed by the drive $W$ as defined in equation \eqref{eq:workwork} equals zero. While the heat dissipated to the environment is given by
\begin{equation}\label{eq:thweak}
\mathcal{Q}=\hbar\omega[\diff N(\omega) -\diff N(\omega)].
\end{equation}

Let us now consider the weak drive limit. The time independent Hamiltonian $H_Q$ is perturbed by the periodic term $\lambda H_D(t)$. To zeroth order in the strength of the drive $\lambda$ we obtain the above expressions for the jump operators and heat. The dressed-qubit only jumps between qubit energy eigenstates exchanging $\pm\hbar \omega$ with the environment. The heat dissipated to the environment is given by \eqref{eq:thweak} is exactly the heat dissipated when the stochastic Schr\"odinger equation is derived in the weak drive limit \cite{Breuer2}. This proves our claim made in the end of Section \ref{sec:ther}.
\section{Monochromatic drive}\label{sec:Mon}
\begin{figure}
\centering
\includegraphics[scale=0.2]{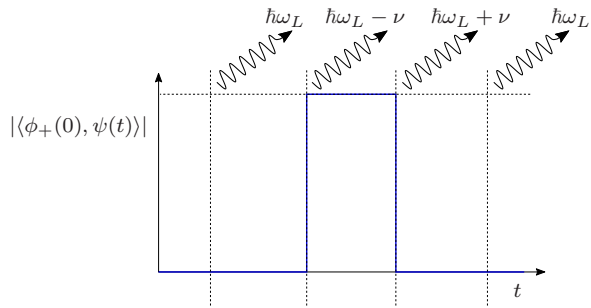}
\setlength{\unitlength}{0.1cm} 
\begin{picture}(0,0)
\put(-35,37.5){\scriptsize$\hbar\omega_L-\nu$}
\put(-21,37.5){\scriptsize$\hbar\omega_L+\nu$}
\put(-5,37.5){\scriptsize$\hbar\omega_L$}
\put(-43,37.5){\scriptsize$\hbar\omega_L$}
\put(-77,23){\scriptsize$|\langle\phi_+(0),\psi(t)\rangle|$}
\put(-10,1){\scriptsize$t$}
\end{picture}
\caption{Possible realisation of the qubit stochastic process. When the photon energy $\omega_L$ is larger than $\nu$, the rate $\gamma(\omega_L-\nu)$ is larger than $\gamma(\nu-\omega_L)$. In this way the qubit gets driven out of the lower Floquet state.}
\label{fig:realisationn}
\end{figure}

Let us now consider the specific example of a monochromatic drive, which is of the form
\begin{equation}
H_d(t)=\lambda(e^{-i\omega_L t}\sigma_++e^{i\omega_L t}\sigma_-).
\end{equation}
We would like to find an expression for the Floquet states to get the effect operators as in equation \eqref{eq:effect2}. We can avoid solving the eigenvalue problem \eqref{eq:floquet} directly by finding a gauge transformation $U_t=e^{iF(t)}$, where $F(t)$ is a Hermitian matrix with the same period as $H_Q(t)$, such that the operator
\begin{equation}\label{eq:gunit}
G=U_t H_Q(t) U^\dagger_t +i\hbar (\partial_t U_t)U^\dagger_t
\end{equation}
is time independent. The Floquet states can now be given in terms of the eigenvectors $|r\rangle$, with eigenvalues $\epsilon_r$ of $G$. It can checked that the states defined as
\begin{equation}\label{eq:ffloquett}
|\phi_r(t)\rangle= U^\dagger_t |r\rangle,
\end{equation}
are indeed eigenstates of the Floquet Hamiltonian with quasi-energies $\epsilon_r$, i.e. they  solve the eigenvalue problem giving in equation \eqref{eq:floquet}.
For the monochromatic drive we can take \cite{GeKoSk95}
\begin{equation}\label{eq:hunit}
F(t)=\omega_L \sigma_+\sigma_-t.
\end{equation}
The Floquet states and quasi-energies can thus be found by diagonalising the matrix
\begin{equation}\label{eq:ggunit}
G=\frac{\hbar}{2}(\omega_q-\omega_L)(\sigma_+ \sigma_- -\sigma_- \sigma_+)+\lambda\left(\sigma_++\sigma_-\right)-\frac{\omega_L}{2}\mathbb{I}.
\end{equation}
The Floquet states and effect operators are explicitly calculated in the appendix. The difference of the quasi-energies is $\epsilon_+-\epsilon_-=\hbar\nu=\sqrt{\hbar^2(\omega_q-\omega_L)^2+4\lambda^2}$. To formulate the thermodynamic quantities for this model it is instructive to have stochastic Schr\"odinger equation for the dressed-qubit
\begin{align}\label{eq:seemon}
\diff \Psi(t)	&=-\frac{i}{\hbar}K(\Psi(t))+\sum_{s=-1,0,1}\bigg[\left(\frac{ B(\omega_L+s\nu)\Psi(t)}{\|B(\omega_L+s\nu)\Psi(t)\|}-\Psi(t)\right)\diff M(\omega_L+s\nu)\\
				&+\left(\frac{B(-\omega_L+s\nu)\Psi(t)}{\|B(-\omega_L+s\nu)\Psi(t)\|}-\Psi(t)\right)\diff M(-\omega_L+s\nu)\bigg].
\end{align}
The operator $K$ has been defined in equation \eqref{eq:ssecontph} and the jump operators $B$ are explicitly defined in Appendix \ref{sec:appMon}.
The dressed-qubit can make six different jumps. They come in three pairs that have the same effect on the qubit but a different amount of energy is exchanged. For example the jumps governed by the Poisson processes $M(\omega_L-\nu)$ and $M(-\omega_L-\nu)$ both bring the qubit to the excited Floquet state. According to our earlier discussions we can interpret the difference in energy as creation or annihilation of photons. In the first jump one extra photon gets annihilated in the drive, in the second jump one photon gets created. Figure \ref{fig:realisationn} shows a possible realisation of the stochastic process for the driven qubit.
The rates satisfy the detailed balance relation \eqref{eq:detbal}.

We can now apply the definitions of Section \ref{sec:ther} to get the thermodynamic quantities for the monochromaticly driven system. The dissipated heat is give by
\begin{equation}\label{eq:heatMon}
\mathcal{Q}=\sum_{s=-1,0,1}(\hbar\omega_L+s\nu)[\diff M(\hbar\omega_L+s\nu)-\diff M(-\hbar\omega_L-s\nu)]
\end{equation}
The thermodynamics of this model have also been studied by \cite{AlickiFloquet}. Their analysis is based on a Markovian master equation of the state operator of a qubit, see \cite{AlickiMaster} and \cite{Szczygielski}. The heat current dissipated to the environment they obtain is the same as the average over all realisations $\mathbb{E}[.]$ of the expression in equation \eqref{eq:heatMon}. We identify the contribution to the heat current as follows. The packets $\hbar \omega_L$ come from the drive. Since the Floquet states are qubit-dressed states, the exchange of $\hbar\nu$ comes from a change in internal energy of the qubit and the drive.

\section{Conclusion}
In this manuscript we considered a periodically driven qubit in contact with a thermal environment. Using Floquet theory we embedded the periodically driven qubit into an infinite dimensional time-autonomous system. This system we called the dressed-qubit. 

We studied the thermodynamics of a periodically driven qubit in contact with a thermal environment by reformulating the thermodynamics of the dressed-qubit. This is where our contribution differs from the existing literature, where the analysis is based on the Lindblad-Gorini-Kossakowski-Sudarshan equation for the state operator in $\mathbb{C}^2$. The advantage of the dressed-qubit is that it is time-autonomous. 

Our approach can be used for the analysis of microscopic or nano-scale devices like \cite{PekCal}. It can be extended to systems with more that two energy levels or multiple heat baths in a straightforward way. It also offers a very clear interpretation in terms of atom-dressed states.

\begin{acknowledgments}
I would like to thank Jukka Pekola, Paolo Muratore-Ginanneschi, Kay Schwieger and Christian Maes for fruitful discussions and comments.

The research has been supported by the Academy of Finland via the Centre of Excellence in Analysis and Dynamics Research (project No. 307333)
\end{acknowledgments}
\appendix

\section{Stochastic Schr\"odinger equation}\label{sec:appSSE}
\subsection{Jump rates}\label{sec:appSSE1}
The jump rates (for both the driven qubit and dressed-qubit) are given by \cite{openpaper, BPfloquet} 
\begin{equation}
\gamma(\omega)=\int_{-\infty}^{+\infty}\diff t e^{i(\hbar\omega +\eta_k-\eta_l) t} \langle\sum_{k,l}g^*_{kl}a^\dagger_k a_l\sum_{p,q}g_{pq}a^\dagger_p a_q)\rangle_\beta
\end{equation}
where the average is over the thermal state of the electron bath with inverse temperature $\beta$. 
Calculating the average and peforming the standard approximation of the sums over different wave numbers as integrals, which is for example explained in Chapter 2 of \cite{solid}, we arrive at
\begin{eqnarray}
\gamma(\omega)	&=\frac{9 N^2\hbar}{4E_F^3}\int_0^\infty \diff E_1\int_0^\infty \diff E_2\int_{-\infty}^{+\infty}\diff t e^{i(\omega +E_1-E_2) t} (1-f(E_1))f(E_2)g^2\sqrt{E_1E_2}
\end{eqnarray}
where $N$ is the average amount of electrons in the bath, $E_F$ is the Fermi energy and $f(x)=\frac{1}{1+e^{\beta(E-\mu)}}$ is the Fermi distribution and we assumed that the coupling $g_{kl}=k$ is constant. Evaluating the time integral gives us a delta function over the energies divided by $2\pi$, evaluating one of the integrals over the energies gives us
\begin{equation}
=\frac{9 N^2\hbar}{4E_F^3(2\pi)}\int_0^\infty \diff E  (1-f(E))f(E+\hbar\omega)\sqrt{E(E+\hbar\omega)}.
\end{equation}
From this one can see that the rates indeed satisfy detailed balance.
\subsection{Embedding}\label{sec:appSSE2}
To show that the relation \eqref{eq:equivvv} indeed embeds the driven qubit into the dressed-qubit we want to show that the short time increment of the dressed-qubit state $\diff\Psi(t,\tau)=\Psi(t+\diff,\tau)-\Psi(t,\tau)$ satisfies the stochastic Schr\"odinger equation for the dressed-qubit \eqref{eq:sseph}. 

First we show that the form \eqref{eq:equivvv} implies that
\begin{equation}\label{eq:probs}
\|A(\omega)\psi(t)\|=\|B(\omega)\Psi(t)\|_{L^2}
\end{equation}
Writing $\psi(t)=\phi_+(0)\langle \phi_+(0),\psi(t)\rangle+\phi_-(0)\langle \phi_-(0),\psi(t)\rangle$ in the Floquet basis, then 
\begin{equation}\label{eq:lift}
\Psi(t,\tau)=e^{i\mu(t)\omega_L \tau}P_{\tau,0}\psi(t)=e^{-i\mu(t)\omega_L\tau}\phi_+(\tau)\langle \phi_+(0),\psi(t)\rangle+e^{-i\mu(t)\omega_L\tau}\phi_-(\tau)\langle \phi_-(0),\psi(t)\rangle. 
\end{equation}
Let us calculate $A(\omega)\psi(t)$ and $B(\omega)\Psi(t)$ using definitions \eqref{eq:effect2} and \eqref{eq:effectph2}.
\begin{equation}
A(\omega)\psi(t)=\sum_{r=\pm} \alpha_{r,+,n_\omega}\phi_r(0)\langle \phi_+(0),\psi(t)\rangle +\alpha_{r,-,n_\omega}\phi_r(0)\langle \phi_-(0),\psi(t)\rangle
\end{equation}
\begin{equation}
B(\omega)\Psi(t)=\sum_{r=\pm} \alpha_{r,+,n_\omega}\phi_{r,-\mu(t)-n_\omega}\langle \phi_+(0),\psi(t)\rangle +\alpha_{r,-,n_\omega}\phi_{r,-\mu(t)-n_\omega}\langle \phi_-(0),\psi(t)\rangle
\end{equation}
From the above equations a straightforward calculation shows that \eqref{eq:probs} indeed holds.
Note that 
\begin{equation}\label{eq:proof1}
(B(\omega)\Psi(t))(\tau)=e^{-i(\mu(t)+n_\omega)\omega_L\tau}P_{\tau,0}A(\omega)\psi(t),
\end{equation} 
and from this it also follows that
\begin{equation}\label{eq:proof2}
(B^\dagger(\omega)B(\omega)\Psi(t))(\tau)=e^{-i\mu(t)\omega_L\tau}P_{\tau,0}A^\dagger(\omega) A(\omega)\psi(t).
\end{equation}

Let us now show that the state $\Psi(t,\tau)=e^{i\mu(t)\omega_L \tau}P_{\tau,0}\psi(t)$ defined as in equation  \eqref{eq:equivvv} evolves by the stochastic Schr\"odinger equation \eqref{eq:sseph} for the dressed-qubit.
We take the time increment on both sides of equation \eqref{eq:equivvv}, following the rules of stochastic calculus see e.g. \cite{Kurt}, we find
\begin{align}
\diff \Psi(t,\tau)	=&e^{-i\mu(t+\diff t)\omega_L\tau}P_{\tau,t_0}\psi(t+\diff t)-e^{-i\mu(t)\omega_L\tau}P_{\tau,t_0}\psi(t)\\
					=&\left( e^{-i\mu(t+\diff t)\omega_L\tau}-e^{-i\mu(t)\omega_L\tau}\right) P_{\tau,t_0}\psi(t)+ e^{-in(t)\omega_L\tau}P_{\tau,t_0}\diff\psi(t)\nonumber\\
					&+\left( e^{-i\mu(t+\diff t)\omega_L\tau}-e^{-i\mu(t)\omega_L\tau}\right) P_{\tau,t_0}\diff\psi(t)
\end{align}
With the definition of $\diff \mu(t)$ \eqref{eq:couple} and the fact that $\diff N(\omega)\diff N(\omega')=\diff N(\omega)\delta_{\omega,\omega'}$ we get
\begin{align}
					=&e^{-i\mu(t)\omega_L\tau} \sum_\omega \diff N(\omega)\left( e^{i\mu_\omega\omega_L\tau}-1\right) P_{\tau,t_0}\psi(t)+ e^{in(t)\omega_L\tau}P_{\tau,t_0}\diff\psi(t)\\
					&+e^{i\mu(t)\omega_L\tau}\sum_\omega\diff N(\omega)\left(  e^{i\mu_\omega\omega_L\tau}-1\right) P_{\tau,t_0}\diff \psi(t)
\end{align}
Using the explicit expression of the stochastic Schr\"odinger equation \eqref{eq:SSE1} and cancelling out terms we arrive to
\begin{align}
= -\frac{i}{\hbar} e^{i\mu(t)\omega_L\tau} P_{\tau,0}G(\psi(t))\diff t+\sum_\omega \left( e^{i\mu(t)\omega_L\tau} e^{in_\omega\omega_L\tau}\frac{P_{\tau,t_0}A(\omega)\psi}{\|A(\omega)\psi(t)\|}-\Psi(t,\tau)\right)\diff N(\omega)
\end{align}

From equations \eqref{eq:probs} and \eqref{eq:proof2} it follows that $e^{i\mu(t)\omega_L\tau} P_{\tau,0}G(\psi(t))=(K(\Psi(t)))(\tau)$. Using \eqref{eq:probs} and \eqref{eq:proof1} we find that 
\begin{equation}
\diff \Psi(t,\tau)= -\frac{i}{\hbar} (K(\Psi(t)))(\tau)\diff t+\sum_\omega \left( \frac{(B(\omega)\Psi(t))(\tau)}{\|B(\omega)\Psi(t)\|_{L_2}}-\Psi(t,\tau)\right)\diff N(\omega)
\end{equation}
This expression is almost the evolution equation for the dressed-qubit \eqref{eq:sseph}. The only difference is that the jumps are governed by different Poisson processes. However since $\|A(\omega)\psi(t)\|=\|B(\omega)\Psi(t)\|_{L^2}$ we see that the conditional expectation values, as defined in equations \eqref{eq:condexpfloq} and \eqref{eq:condexpph}, are equal $\mathbb{E}(\diff N(\omega)|\psi(t))=\mathbb{E}(\diff M(\omega)|\Psi(t))$ at all times. This proves that the above stochastic process is equal to the stochastic Schr\"odinger equation \eqref{eq:sseph} for the dressed-qubit.

\section{Monochromatic drive}\label{sec:appMon}
The Floquet can be found by diagonalising \eqref{eq:ggunit} and using equation \eqref{eq:ffloquett}. We get
\begin{align}
\phi_+(\tau)&=(\cos(\theta/2)e^{-i\omega_L \tau}, \sin(\theta/2))\\
\phi_+(\tau)&=(-\sin(\theta/2)e^{-i\omega_L \tau}, \cos(\theta/2)),
\end{align}
where $\cos\theta=\frac{\omega_q-\omega_L}{\sqrt{\hbar^2(\omega_q-\omega_L)^2+4\lambda^2}}$. The difference in the quasi-energies is $\epsilon_+-\epsilon_-=\hbar\nu=\sqrt{(\omega_q-\omega_L)^2+4\lambda^2}$.
The matrix elements of $(\sigma_++\sigma_-)$ in the Floquet basis are
\begin{align}
\langle \phi_+(\tau),(\sigma_++\sigma_-)\phi_+(\tau)\rangle&=e^{-i\omega_L \tau
}\frac{\sin\theta}{2}+ e^{i\omega_L \tau}\frac{\sin\theta}{2}\\
\langle \phi_-(\tau),(\sigma_++\sigma_-)\phi_-(\tau)\rangle&=-e^{-i\omega_L \tau}\frac{\sin\theta}{2}-e^{i\omega_L \tau}\frac{\sin\theta}{2}\\
\langle \phi_+(\tau),(\sigma_++\sigma_-)\phi_-(\tau)\rangle&=-e^{-i\omega_L \tau}\sin^2(\theta/2)+e^{i\omega_L \tau}\cos^2(\theta/2)\\
\langle \phi_-(\tau),(\sigma_++\sigma_-)\phi_+(\tau)\rangle&=e^{-i\omega_L \tau}\cos^2(\theta/2)-e^{i\omega_L \tau}\sin^2(\theta/2).
\end{align}
From the above matrix elements we can see that there are six jump operators, acting on a state $\Psi$ as
\begin{align}
(B(\omega_L)\Psi)(\tau)&=\frac{\sin\theta}{2}\sum_n(\phi_{+,n-1}(\tau)\langle\phi_{+,n},\Psi\rangle_{L^2}-\phi_{-,n-1}(\tau)\langle\phi_{-,n},\Psi\rangle_{L^2})\\
(B(\nu+\omega_L)\Psi)(\tau)&=\sin^2(\theta/2)\sum_n\phi_{-,n-1}(\tau)\langle\phi_{+,n},\Psi\rangle_{L^2}\\
(B(\nu-\omega_L)\Psi)(\tau)&=\cos^2(\theta/2)\sum_n\phi_{-,n-1}(\tau)\langle\phi_{+,n},\Psi\rangle_{L^2}
\end{align}
and their complex conjugates. From equation \eqref{eq:condexpfloq}  we can see that it is possible to get rid of all the sines and consines in the definition of the jump operators by redefining the jump rates as the original rates times the prefactor of the corresponding jump operator squared (for the jumps themselves the prefactors do not matter, the jumps are normalised). For example we define $\frac{\sin^2\theta}{4}\gamma(\omega_L)\rightarrow \gamma(\omega_L)$. We do this for notational simplicity.

\bibliography{lit}

\end{document}